# Optimized Body Deformation in Dragonfly Maneuvers

Ayodeji T. Bode-Oke[1], Samane Zeyghami[2], Haibo Dong [3]
*Department of Mechanical and Aerospace Engineering, University of Virginia, Charlottesville, VA, 22904*

**Abstract**

**Tail bending associated with maneuvering flight of insects is a known phenomenon although there are only a few studies which analyze and quantify the effects and benefits of body configuration changes. We hypothesized that these configuration changes help reduce the energy expenditure in flight. This is quantified by the magnitude of the aerodynamic torque generated by the insect during the maneuver. To test our hypothesis, a dragonfly body model was constructed with the ability to bend at the joint between the thorax and the tail. An optimization problem was defined to find the dynamic body configuration which minimizes the total pitch and yaw torque. The magnitude of the tail deflection was found to be directly correlated with the yaw velocity of the body. Most importantly, our results indicate that for executing the same aerial maneuver, an insect with a flexible body was found to require substantially smaller flight torque when compared to an insect with identical morphology but possessing a rigid body. In addition, changes in the instantaneous mass distribution of the body had the most substantial effect on reducing the flight torque, while the inertial term due to the tail movement had a smaller effect.**

## Nomenclature

| | | |
|---|---|---|
| $I$ | = | moment of inertia |
| $m_i$ | = | mass of each slice of dragonfly body |
| M | = | body mass of real dragonfly |
| $l_i$ | = | length of equivalent cylinder representing each slice of dragonfly body |
| L | = | body length of real dragonfly |
| $p$ | = | roll velocity |
| $q$ | = | pitch velocity |
| $r$ | = | yaw velocity |
| $r_i$ | = | radius of equivalent cylinder representing each slice of dragonfly body |
| $\vec{\tau}$ | = | torque |
| $\vec{\omega}$ | = | body angular velocity vector comprising of body angular velocities |

## I. Introduction

To change flight heading, insects perform a variety of aerial maneuvers. In many maneuvering flights such as predatory or evasive flights, the dynamics of the motion is as important as the final change in the flight orientation. This may be due to several reasons, such as a desired flight path or considerations like the energetic cost of the flight. Therefore, in such situations, the required flight torque is dictated by the body motion as well as its morphology such as mass and moment of inertia neglecting the limitations due to the insects control system and sensory equipment etc. If the body is rigid, the only source of torque generation is the wing which generates aerodynamic force by moving back and forth through the air. Changes in the wing kinematics are the most important means by which an insect alters the flight[1]. However, if the body is capable of deforming as shown in Fig. 1, changes in the mass distribution can affect the motion by two means; first by generating inertial torque associated with the rate of change of the body moment of inertia and second by changing the response of the body to the flight torque by varying the instantaneous distribution of the mass. In the majority of studies on insect flight dynamics, the rigid body assumption is used in study to simplify the analysis[2,3]. However, this assumption imposes several restrictions on the problem such as aforementioned. Although the existence of abdominal deformation, i.e. flexion and deflection, in insects such as locusts and fruit flies has been known for decades[4-6], the extent to which the flexibility of the body affects the flight is not clear yet. Scientists and engineers have inferred from observations that insects may bend their bodies to decrease the moment of inertia about the axis of rotation and to dampen the effects of perturbations for stability purposes using their abdomen as a control surfaces[7]. Abdominal deflection is also employed as a strategy to create asymmetric body drag profiles to increase or decrease rotational inertia. In essence, maneuvers are not restricted only to asymmetric wing kinematics changes[8] but they employ a more complicated approach to increase efficacy and efficiency of their flight.

---

[1] Undergraduate Student, atb5dc@virginia.edu, Student Member.
[2] Graduate Student, sz3ah@virginia.edu, Student Member.
[3] Associate Professor, hd6q@virginia.edu, AIAA Associate Fellow.

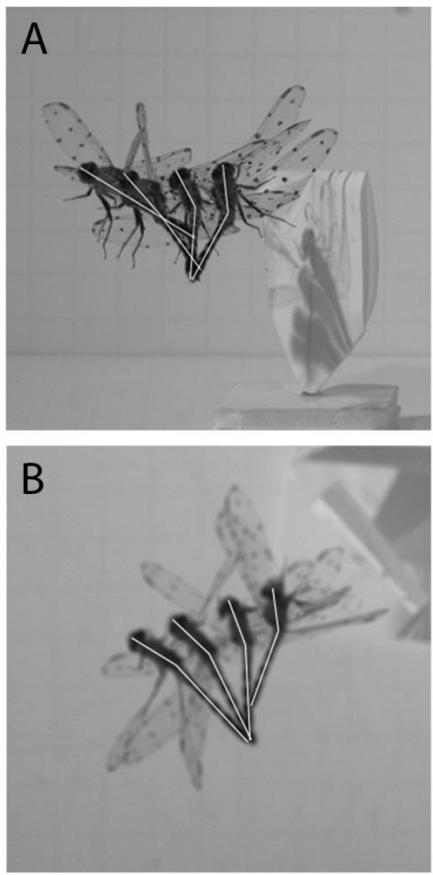

**Figure 1. High Speed Photograph Indicating Tail Bending in Dragonflies.** A selected sequence of images of a dragonfly in free flight is shown. (A) and (B) are images taken by the forward facing and downward facing cameras, respectively. The orientations of the thorax and the tail are indicated by white lines for clarity.

In this study, we focus on how body deformation affects the dynamics of motion. To enumerate the effects of moment of inertia changes due to body deformation on the dynamics of flight during a maneuver, we built a flexible body model and developed computer codes to accurately calculate the moment of inertia while taking into account body flexibility effects. To understand the physics behind body deflection, we hypothesized that the tail of a dragonfly deflects dynamically to minimize the average total torque during a maneuver. Here, we developed an optimization problem to carry out the investigation.

## II. Materials and Methods

### A. Calculation of Moment of Inertia

Moment of inertia (MOI) is a measure of the resistance of a body to rotation and because insect body morphologies are not comprised of regular geometric shapes, their body shapes are approximated in order to calculate the MOI[9-13]. On the contrary, in this paper, we accurately calculate the moment of inertia tensor with the confidence of reliability in all degrees of freedom. The moment of inertia for the dragonfly was obtained from a 3D model constructed in MAYA (Autodesk, San Rafael, CA, USA) based on images of a common species; *Erythimus simplicicollis*, found in North America. Two images; a top and side view are necessary for accurately constructing the 3D model. The non-dimensional body morphological data obtained from the dragonfly model is included in Table 1. The MOI data is non-dimensionalized in Table 1 by $M/L^2$, where M is the body mass (252.52 mg) and L is the body length (43.95 mm).

For analysis, the 3D model was sectioned into 21 pieces as shown in Fig. 2. Slices 1-3, 4-7 and 8-21 make up the head, thorax and tail regions respectively. We established that 21 slices provided enough accuracy for our study. However, since a computer algorithm is responsible for the calculation of MOI, the number of sections can be infinitely increased and the thickness of each section need not be uniform.

**Table 1**: **Geometric Data of each slice of the dragonfly body.** The non-dimensional length and volume of each slice of the dragonfly model are recorded. The non-dimensional MOI included in the table are non-dimensionalized by the body mass and length squared.

|  | Section # | $\ell_i$ | $V_i$ | $I_{xx}^* \times 10^4$ | $I_{yy}^* \times 10^4$ | $I_{zz}^* \times 10^4$ | $I_{xz}^* \times 10^4$ |
|---|---|---|---|---|---|---|---|
| Head | 1 | 3.93E-02 | 2.00E-04 | | | | |
| | 2 | 5.71E-02 | 6.24E-04 | | | | |
| | 3 | 6.81E-03 | 3.69E-05 | | | | |
| Thorax | 4 | 3.52E-02 | 2.76E-04 | 15.00 | 97.00 | 96.00 | -0.44 |
| | 5 | 5.76E-02 | 7.79E-04 | | | | |
| | 6 | 6.63E-02 | 9.03E-04 | | | | |
| | 7 | 8.66E-02 | 6.85E-04 | | | | |
| Inner Tail | 8 | 4.27E-02 | 2.46E-04 | | | | |
| | 9 | 6.59E-02 | 4.10E-04 | | | | |
| | 10 | 3.91E-02 | 1.40E-04 | | | | |
| Outer Tail | 11 | 3.76E-02 | 7.71E-05 | | | | |
| | 12 | 4.19E-02 | 6.70E-05 | | | | |
| | 13 | 5.02E-02 | 6.00E-05 | | | | |
| | 14 | 5.86E-02 | 8.12E-05 | 1.50 | 230.80 | 230.50 | -6.80 |
| | 15 | 4.93E-02 | 6.46E-05 | | | | |
| | 16 | 5.11E-02 | 9.45E-05 | | | | |
| | 17 | 5.40E-02 | 1.01E-04 | | | | |
| | 18 | 3.43E-02 | 7.16E-05 | | | | |
| | 19 | 5.89E-02 | 1.16E-04 | | | | |
| | 20 | 3.42E-02 | 5.05E-05 | | | | |
| | 21 | 3.32E-02 | 1.91E-05 | | | | |
| | **Total** | **1.00E+00** | **5.10E-03** | **16.5** | **327.8** | **326.5** | **-7.24** |

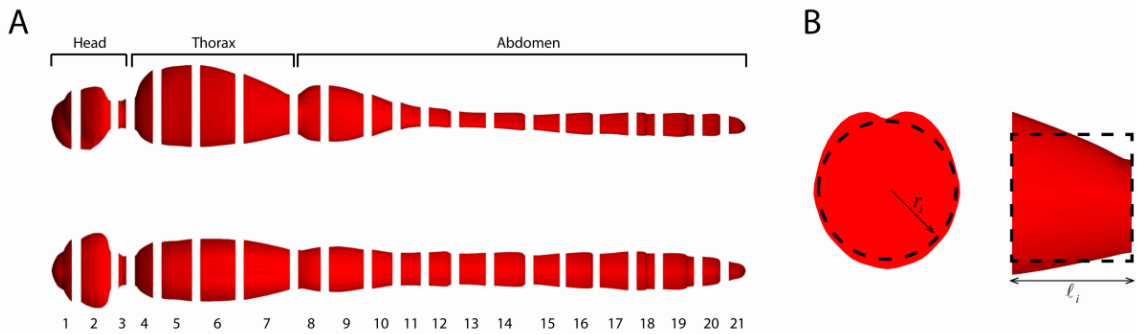

**Figure 2. Dragonfly model for MOI calculations**. Figure A depicts a sliced dragonfly model. The model is sliced into 21 pieces. Slices 1-3, 4-7, 8-21 make up the head, thorax, and abdomen, respectively. Figure B shows the approximation of an $i^{th}$ slice as a cylinder having a volume equivalent to the slice. $r_i$ shows the corresponding radius of the approximated cylinder and $l_i$ shows the length of the cylinder. The length of the cylinder is equal to the length of each slice.

From the 3D model, we obtained the non-dimensional geometric center of mass as well as the volume of each slice. Afterward, each individual slice was approximated as a cylinder with a volume equivalent to the actual slice volume to quantify the contribution of the slices to the MOI. The equivalent cylinder has the same length as the slice and the circular cross-sectional area is oriented in the direction of the roll axis. The mass of each slice is computed by sectioning a real dragonfly into three pieces; head, thorax and abdomen. We assumed that density is constant across each section. The real insect body length is used to dimensionalize all the length quantities. From this dimensionalizaion, we obtain the dimensional volume of each slice and then multiply density by volume to obtain the mass of each slice. The tensor of MOI of the dragonfly was calculated by adding the MOI of each slice about the axes fixed at the center of mass of the body. $I'_{head}, I'_{thorax}, I'_{tail}$ are tensors of moment of inertia of the slices of the head, thorax, tail, respectively, about the axes originating from their own center of mass and fixed to them. $I_{trans_i}$ represents the tensor which contains the parallel axis theorem that enables us to calculate the effect of distance from the body center of mass to the center of mass of each individual slice.

$$I = \sum_{i \in n} I'_i + I_{trans_i} \qquad n = \{head, thorax, \text{abdomen}\}$$

$$I'_i = \frac{m_i}{12} \begin{bmatrix} 6r_i^2 & 0 & 0 \\ 0 & 3r_i^2 + l_i^2 & 0 \\ 0 & 0 & 3r_i^2 + l_i^2 \end{bmatrix} \qquad I_{trans_i} = m_i \begin{bmatrix} y_i^2 + z_i^2 & x_i y_i & x_i z_i \\ y_i x_i & x_i^2 + z_i^2 & y_i z_i \\ z_i x_i & z_i y_i & x_i^2 + y_i^2 \end{bmatrix}$$

**B. Optimization Problem Definition**

Our objective in defining the optimization problem was to find out whether body bending patterns during maneuvering fight of dragonflies are beneficial to the dynamics of the system or are rather passive reactions. In other words, if the system is free to stay rigid or deflect during the maneuver will it choose to deflect and if so how this deflection will change during the course of the maneuver or how would it benefit the insect? To investigate, we defined an optimization problem with the objective of minimizing the average total flight torque needed to perform a specific maneuver. We modeled the tail as a free bidimensional pendulum so that the dragonfly can continuously deflect its tail in any direction at any time and change it during the course of the maneuver, if desired. We used a 4-control point B-spline curve to represent time history of angle of rotation. Similarly, each of three components of the axis of rotation of the tail is expressed by its respective B-spline curve. Each B-spline has two fixed and two variable control points and optimization algorithm is able to locate four pair of control points (corresponding to four B-splines) which finally define time history of change in axis and angle of rotation of the tail. Both the schematic for the B-spline curves as well as the optimization flow chat is show in Fig 3. For a known maneuver, the instantaneous magnitude of the total flight torque can be calculated as follows:

$$\vec{\tau}_{total} = I\vec{\dot{\omega}} + \vec{\omega} \times I\vec{\omega} + \dot{I}\vec{\omega} \qquad (1)$$

Where $\tau_{total}$ is the vector of flight torque. The first two terms on the right hand side of Eqn. (1) comprise the aerodynamic torque and the last term is an inertial torque due to the tail deflection. Note that that so called aerodynamic torque in this study, includes some inertial terms due to coupling between the three rotational motions but an investigation into the coupling is beyond the scope of this work. A detailed discussion on those inertial terms due to coupling and their effect on the dynamics of flight can be found in Ref. 10.

The cost function is directly correlated with energy expenditure in flight may be defined to minimize the average or the maximum flight torque. The cost function chosen in this study is calculated as,

$$\cos t = \left| mean(\tau_{pitch}) + mean(\tau_{yaw}) \right| \qquad (2)$$

Where $\tau_{pitch}$ and $\tau_{yaw}$ are the aerodynamic components of the flight torque in yaw and pitch directions. The optimization was performed using a built-in optimizer (fmincon) in MATLAB (Mathworks, Natick, MA, USA). We chose "interior point" as the search algorithm.

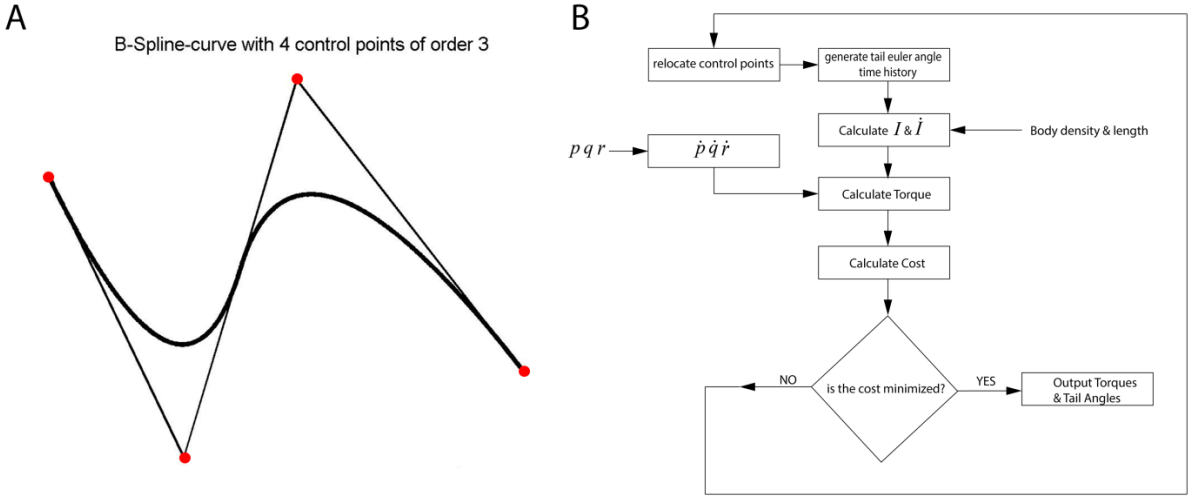

**Figure 3. Optimization Problem Definition**. Figure 3A shows a third order B-spline curve with 4 control points. The control points are depicted as red circles. The B-spline curve help generate the time history of the tail euler angles. Figure 2B represents the optimization flow chart. The inputs into the optimization algorithm are body density and length as well as body angular velocities, pqr. The volume of each slice of the dragonfly is obtained by dimensionalizing the non-dimensional volume by body length. Having the mass distribution, MOI and time rate of change of MOI can be calculated. The time rate of change of body angular velocities can be calculated also. The torque is calculated given angular velocities, rate of change of MOI, and rate of change of MOI. The cost function then calculates the energy expenditure. If the energy expenditure is minimal, the optimization algorithm output the tail angles and torques. Otherwise, the optimization is iterated by relocating the control points on the B-spline curves and the generating new tail euler angles. The process iterates till minimal energy expenditure is attained.

## III. Results and Discussion

### A. Effect of Tail Deflection on Changes in MOI

As aforementioned in the introduction, by deflecting the tail, the dragonfly alters the moment of inertia tensor. In order to quantify effect of the tail deflection on the MOI tensor changes, we varied the tail pitch angle between -45 deg to 45 deg while keeping the tail yaw angle constant and vice versa. The ranges of tail deflections are chosen based on free flight observations of dragonflies. In both cases, we observed that by deflecting the tail, pitchwise or yawwise, the moment of inertia around the yaw and pitch axis can be enhanced up to twice its original value. At erect posture, the roll moment of inertia of a dragonfly body is about 2 orders of magnitude smaller than the yaw or pitch MOI. This suggests that the response of the dragonfly body to the roll torque is significantly faster than that of the pitch or yaw. This may cause discrepancies in the sensitivity of the insect motion sensors. By deflecting the tail, Ixx can be increased more than 20 times its original value which may be necessary for controlling the roll motion (Fig. 4A and B). We also observed that the products of inertia were also influenced by yawing and pitching the tail. The products of inertia are measures of body symmetry . When the tail is deflected pitchwise, there is an increase in asymmetry in the xz plane. Likewise when the yawwise deflection occurs, there is asymmetry in the xy plane. The results of our analysis in this section indicate that the variations in the MOI tensor can be potentially significant during the flight.

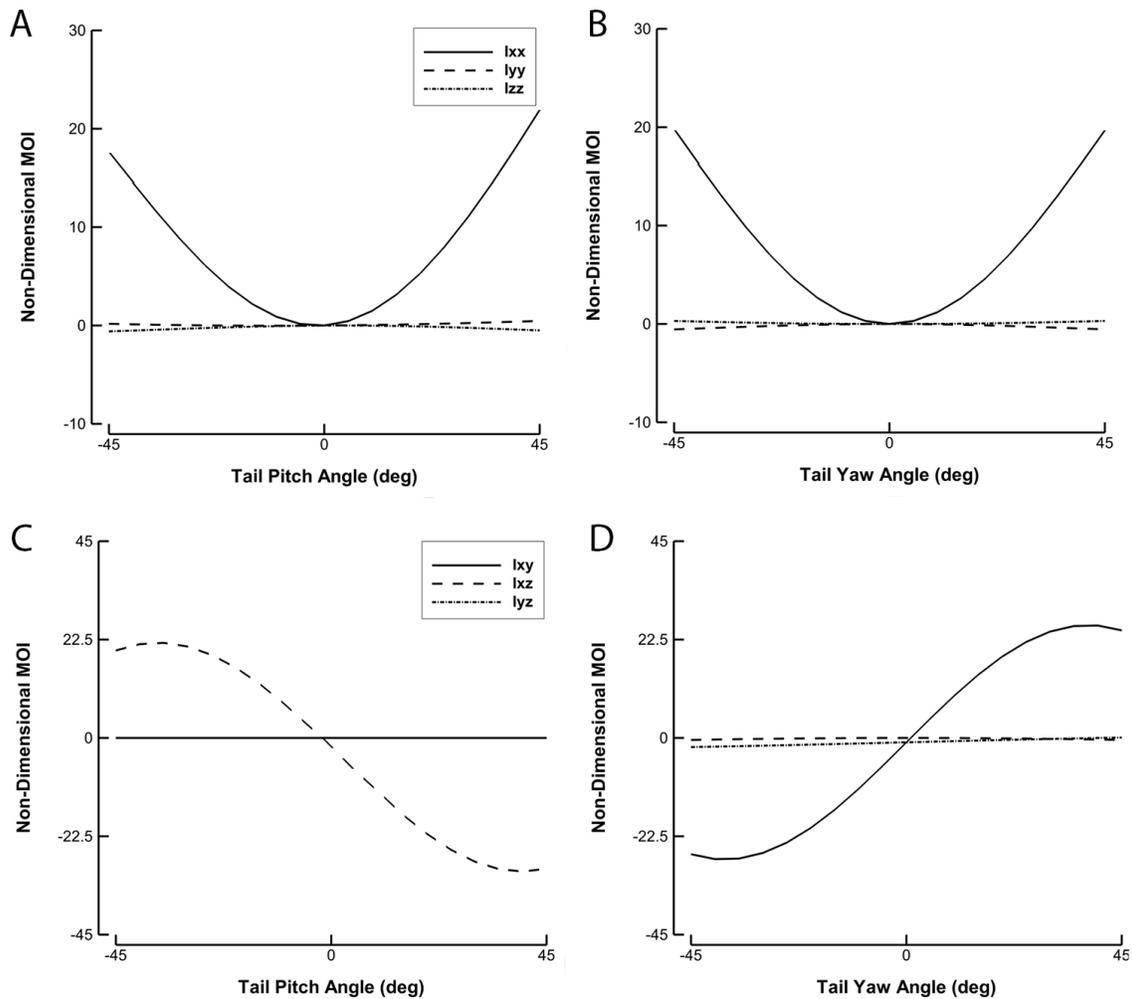

**Figure 4. Effect of variation of tail euler angles on MOI.** The figures (A, B, C and D) show the amount of change that takes place in the moment of inertia when the tail is pitched only or yawed only. The tail angles are varied between -45 to 45 degrees. 0 degrees represents the erect position of the dragonfly. The top two figures, A and B, illustrate that when the moment of inertia is varied in pitch and yaw axis, the greatest variation occurs in the rolling moment of inertia. The rolling moment of inertia was about 20 times its original value while the pitching and yawing MOI did not change much. The bottom figures, C and D, also illustrate the change in body symmetry when the tail is bent pitch wise or yaw wise.

### B. Optimized Tail Deflection and its Relationship to Body Yaw Velocity.

To investigate the effect of tail bending on the overall torque generation in dragonfly maneuvering flight, we first used the real Euler angles of a dragonfly in a turning takeoff maneuver and calculated the optimized posture of the tail during that flight. A turning takeoff is a flight maneuver during which a dragonfly changes the flight heading by more than 90 degrees while the center of mass is elevated over several body lengths[10, 14]. The body motion reconstruction is performed using an accurate method which is described comprehensively in Ref. 15. Figure 5A, shows the body angular velocities, the optimized tail angles and the total torque (thin solid line), aerodynamic torque (dashed line) and rigid body torque (thick solid line). The rigid body torque is calculated by balancing the dynamics of the motion using the known values of body angular velocities and accelerations as well the moment of inertia of the rigid and erect dragonfly body. The acceleration and deceleration phase of the maneuver is respectively defined as the duration before and after the maximum body yaw velocity was attained. The acceleration phase is shaded in Fig. 5 for clarity. During the maneuver, both the yawwise and pitchwise deflections of the tail increases as the body yaw velocity increased. The maximum deflection in the tail yaw angle was reached at the point that the body yaw velocity was maximum. The maximum pitchwise deflection of the tail happened earlier during the maneuver, similar to the occurrence of the maximum pitch velocity of the body. One important observation is that the total torque required by a rigid body is substantially different from that of a flexible body with an optimized posture. Note that if the body is rigid, there is only one single solution to Eqn. (1), meaning that if $\vec{\omega}$ and consequently $\dot{\vec{\omega}}$ are known, there is only one solution for torque. However, for a flexible body there are infinite ways of executing the same motion. Body flexibility can change the flight torque by two means. First, by offering inertial torque terms due to the tail deflection; $\dot{I}\vec{\omega}$, and second, by manipulating the instantaneous magnitude of each element in the tensor of MOI. Comparing the yaw torque for a dragonfly with a rigid body (thick solid blue line) with that of one with a flexible body (thin solid blue line), we can infer that that the yaw toque of the flexible body is significantly lower than that of a rigid body, especially during the acceleration phase. Similarly, the pitch torque of a rigid body stays high throughout the maneuver while that of the flexible body drops to substantially smaller magnitudes during early stages of the flight. The roll torque of the flexible body with the optimized body posture is larger than that of the rigid body. That is due to our choice of cost function in this study which assumes the cost of rolling the body is significantly smaller than that of yawing or pitching. The cost function can be modified to incorporate a cost to generate roll torque as well. The difference between the thin solid line and dashed line in Fig. 5 shows the contribution of the inertial torque term due to the tail deflection; the last term on the right hand side of Eqn. (1). It is evident from Fig. 5 that inertial term is close in magnitude to the aerodynamic torque. Note that as the tail bends, it does not always generate inertial torque that is in favor of the maneuver. In fact, during the acceleration phase, the inertial torque due to the tail deflection decreased the total torque. In interpreting these results one needs to remember, as was previously mentioned, that this inertial tem is not the only contribution of the body flexibility to the flight torque and the overall benefit of the tail bending can only be understood by considering all three kinds of lines; thick solid, thin solid and dashed, that are plotted in torque graph. Furthermore, the maximum deflection of the tail is only about 15 degrees yawwise and 10 degrees pitchwise. These results imply that small abdominal deflections during maneuvering flight are enough to enhance the flight performance.

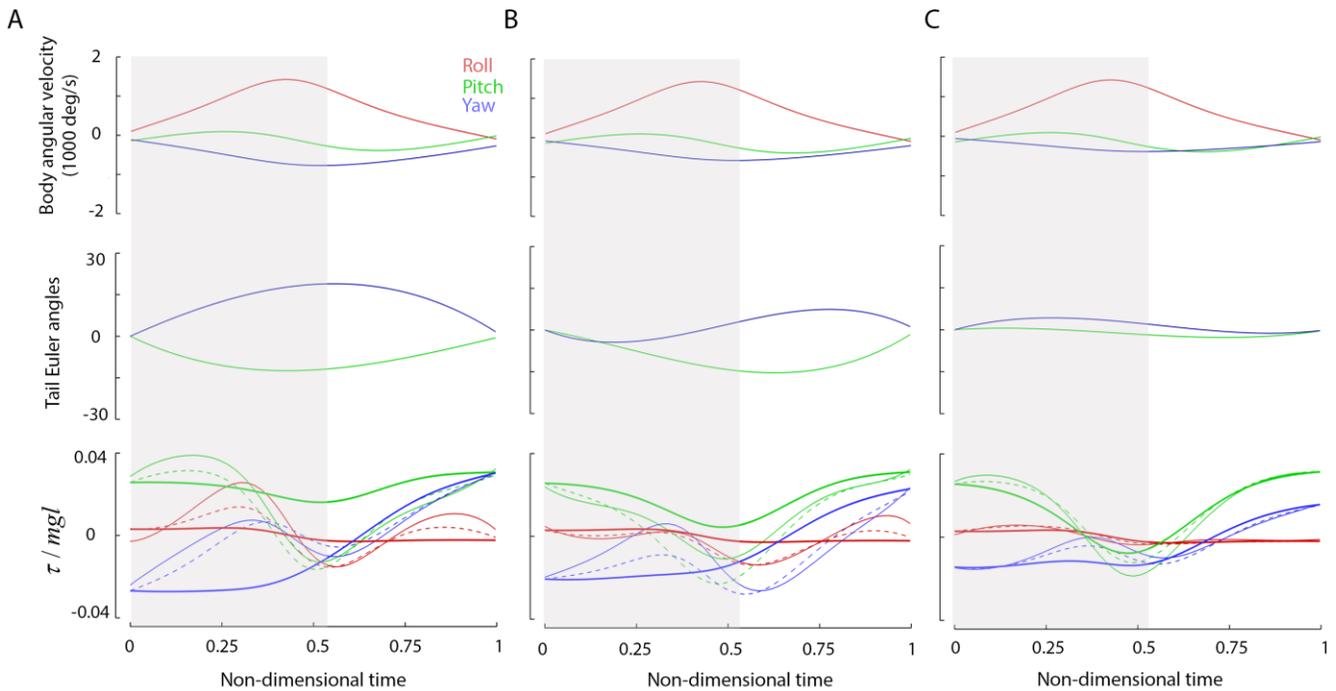

**Figure 5. Optimized tail bending kinematics and the resultant flight torque.** This figure (A, B and C) comprises of body angular velocity data, the optimized tail angles as well as the torques obtained from the cost function. Figure A, represent optimized results based on real insect flight data while B and C represent data obtained when the body yaw velocity was decreased to 75% and then 50%, respectively. In the torque plots shown on the third row, the thick solid lines represent the rigid body torques, the thin dashed lines represent the aerodynamic torque and the thin solid lines represent the total flight torque.

To investigate the connection between the optimized body posture and the flight, we systematically altered the body yaw velocity and solved for the optimized tail deflection. The body yaw velocity was decreased to 75% and then 50% and the results are shown in Fig. 5B and C. Visual inspection of Fig. 5 immediately indicates that the magnitude of the tail deflection is directly correlated with the magnitude of the body yaw velocity; meaning that more deflections are required for faster maneuvers. The other important observation is that both tail yaw and pitch angles vary accordingly with the yaw velocity even though the pitch velocity is identical for all three cases represented in Fig. 5. In all three cases in Fig. 5, the magnitude of the inertial torque due to the tail bending is smaller in acceleration phase when compared to deceleration phase.

## IV. Conclusion

Taken together, our results imply that body flexibility benefits the flight performance by offering more ways by which a flight can be achieved. This is accomplished by controlling the instantaneous mass distribution of the body as well as the generating inertial torques due to movement of the tail with respect to the thorax. The former changes the response of the body to the generated torque and enhances or decreases the resistance of the body to a specific motion. The latter acts by inserting inertial terms that are proportional to the rate of change in the mass distribution. Our results indicates that both these effects play an essential role in reducing the average flight torque required for the flight with the former effect being more dominant. Having tackled the tail deflection problem comprehensively, there are still many questions to be investigated. In the future, we hope to investigate the influence of body shape on the optimized body posture. For example, insects with round and stocky bodies are less commonly observed to deflect their abdomen during flight. Although, this may be related to the body morphological limitations, we expect that the body geometry as well as the mass distribution would influence the optimized results.

## Acknowledgments

This work was partially supported by NSF [grant number CEBT-1313217].

## References

[1]Dudley, R. *The biomechanics of insect flight: form, function, evolution*. Princeton: Princeton University Press, 2002.

[2]A. J. Bergou, L. R., J. Guckenheimer, I. Cohen, Z.J. Wang. "Fruit Fly Modulate Passive Wing Pitching to Generate In-Flight Turn," *Phys. Rev. Lett.* Vol. 104, 2010.

[3]Cheng, B., Fry, S., Huang, Q., Deng, X. "Dynamics and Control of Turning in Fruit Fly Drosophila," *Journal of Experimental Biology*, 2009, p. In press.

[4]Camhi, J. M. "Sensory control of abdomen posture in flying locusts," *J. exp. Biol* Vol. 52, No. 3, 1970, p. 533.

[5]Götz, K. G., Hengstenberg, B., and Biesinger, R. "Optomotor control of wing beat and body posture in Drosophila," *Biological Cybernetics* Vol. 35, No. 2, 1979, pp. 101-112.

[6]Hinterwirth, A. J., and Daniel, T. L. "Antennae in the hawkmoth Manduca sexta (Lepidoptera, Sphingidae) mediate abdominal flexion in response to mechanical stimuli," *Journal of Comparative Physiology A* Vol. 196, No. 12, 2010, pp. 947-956.

[7]Dyhr, J. P., Morgansen, K. A., Daniel, T. L., and Cowan, N. J. "Flexible strategies for flight control: an active role for the abdomen," *The Journal of experimental biology* Vol. 216, No. 9, 2013, pp. 1523-1536.

[8]Finio, B. M., and Wood, R. J. "Distributed power and control actuation in the thoracic mechanics of a robotic insect," *Bioinspiration & biomimetics* Vol. 5, No. 4, 2010, p. 045006.

[9]Zeyghami, S., and Dong, H. "A Body Reorientation Strategy in Insect Takeoff Flight," *50th AIAA Aerospace Sciences Meeting*. 2012.

[10]Zeyghami, S., and Dong, H. "Study of turning takeoff maneuver in free-flying dragonflies: effect of dynamic coupling," *arXiv preprint arXiv:1502.06858*, 2015.

[11]Combes, S. A., and Dudley, R. "Turbulence-driven instabilities limit insect flight performance," *Proceedings of the National Academy of Sciences* Vol. 106, No. 22, 2009, pp. 9105-9108.
doi: 10.1073/pnas.0902186106

[12]Hedrick, T. L., and Daniel, T. L. "Flight control in the hawkmoth Manduca sexta: the inverse problem of hovering," *Journal of Experimental Biology* Vol. 209, No. 16, 2006, pp. 3114-3130.

[13]Taylor, G. K., and Thomas, A. L. R. "Dynamic flight stability in the desert locust Schistocerca gregaria," *Journal of Experimental Biology* Vol. 206, No. 16, 2003, pp. 2803-2829.

[14]Zeyghami, S., and Dong, H. "A Body Reorientation Strategy in Insect Takeoff Flight," 2012.

[15]C. Koehler, Z. L., Z. Gaston, H. Wan, H. Dong. "3D reconstruction and analysis of wing deformation in free-flying dragonflies," *Journal of Experimental Biology*, 2012.